\documentclass[
  aps,
  prb,
  twoside,
  twocolumn,
  showpacs,
  floatfix,
  10pt,
  superscriptaddress,
]{revtex4-1}

\usepackage{graphicx}
\usepackage{amsmath}
\usepackage{color}

\newcommand{\Eq}[1]{Eq.~(\ref{#1})}
\newcommand{\eq}[1]{(\ref{#1})}

\newcommand{\fig}[1]{Fig.~\ref{#1}}

\newcommand{\sect}[1]{Sect.~\ref{#1}}

\renewcommand{\vec}[1]{\ensuremath\boldsymbol #1}
\newcommand{\VCSGC}{\text{V}}
\newcommand{\SGC}{\text{S}}
\newcommand{\C}{\text{C}}
\newcommand{\K}{\text{K}}
\newcommand{\dmu}{\ensuremath{\Delta\mu}}

\begin{document}

\title{
  Calculations of excess free energies of precipitates \\
  via direct thermodynamic integration across phase boundaries 
}

\author{Babak Sadigh}
\email{sadigh1@llnl.gov}
\affiliation{
  Lawrence Livermore National Laboratory,
  Condensed Matter and Materials Division,
  Livermore, California, 94551, USA
}
\author{Paul Erhart}
\email{erhart@chalmers.se}
\affiliation{
  Lawrence Livermore National Laboratory,
  Condensed Matter and Materials Division,
  Livermore, California, 94551, USA
}
\affiliation{
  Chalmers University of Technology,
  Department of Applied Physics,
  Gothenburg, Sweden
}

\begin{abstract}
We describe a technique for constraining macroscopic fluctuations in thermodynamic variables well-suited for Monte Carlo (MC) simulations of multiphase equilibria. In particular for multicomponent systems this amounts to a statistical ensemble that implements constraints on both the average composition as well as its fluctuations. The variance-constrained semi-grandcanonical (VC-SGC) ensemble allows for MC simulations, in which single-phase systems can be reversibly switched into multiphase equilibria allowing the calculation of excess free energies of precipitates of complex shapes by thermodynamic integration. The basic features as well as the scaling and convergence properties of this technique are demonstrated by application to an Ising model. Finally, the VC-SGC MC simulation technique is used to calculate $\alpha/\alpha'$ interface free energies in Fe--Cr alloys as a function of orientation and temperature taking into account configurational, vibrational, and structural degrees of freedom.
\end{abstract}

\date{\today}


\maketitle


\section{Introduction}

Multiphase systems are ubiquitous in nature. Since their properties are crucially dependent on the interfaces between the phases involved interface free energies constitute crucial input to kinetic and phase field models. \cite{BroLeFin08} Their calculation, however, often constitutes a complicated task. \cite{BroGil86c, DavLai03, HoyAstKar01, MisAstLi10, AngCerLee10} Here, an extension of conventional thermostatistical ensembles. Is described that enables calculation of excess free energies via thermodynamic integration across phase boundaries. The approach presented in this paper can be viewed as a generalization of the extended Gaussian ensemble technique to multicomponent systems. \cite{Het87,  JohPlaViv03}

An equilibrium ensemble in statistical mechanics is a collection of macroscopic systems that have been prepared in the same thermodynamic state. This can be accomplished by constraining one of each pair of conjugate thermodynamic variables that represent generalized displacements and forces, e.g. volume--pressure, entropy--temperature, or particle number--chemical potential. In single-phase regions of the phase diagram, there is a one-to-one correspondence between generalized displacements and forces. This allows for straightforward computation of free energy differences by thermodynamic integration. In contrast, in two-phase regions of the phase diagram, the mapping from generalized forces to generalized displacements becomes multi-valued. For instance the melting temperature maps to two distinct entropies, namely those of the solid $S_s$ and the liquid $S_l$, respectively.  Multiphase equilibria can be obtained by constraining the entropy to values $S_s<S<S_l$. If the conjugate forces are observables of the ensemble this enables thermodynamic integration across phase boundaries. Conveniently, two important conjugate forces, namely temperature and pressure, are readily available in the microcanonical ($NVE$) ensemble. In contrast the chemical potential is not an observable of ensembles that control the number of particles or chemical composition in multicomponent systems. 

In this paper a technique is described for constructing ensembles, which constrain fluctuations of generalized displacements, that can be viewed as an extension and generalization of the Gaussian ensemble technique. \cite{Het87, JohPlaViv03} Such {\em variance-constrained} (VC) ensembles provide direct access to multiphase regions of the phase diagram while allowing for observation of both generalized forces and displacements. Thereby they enable thermodynamic integration across phase boundaries and computation of free energies for compositions inside miscibility gaps. In this paper this technique is employed to study precipitation in an Ising model system and a binary Fe--Cr alloy. Through Monte Carlo simulations in the VC semi-grandcanonical (SGC) ensemble chemical potential--concentration isotherms in the shape of the van-der-Waals loop are obtained. In this fashion, one can extract interface free energies as a function of temperature and orientation. It is also demonstrated that interface free energies can be calculated by monitoring the growth and/or shrinkage of compact precipitates.

The paper is organized as follows. After recapitulating key features of the related canonical and semi-grandcanonical ensembles, the next section introduces the VC-SGC ensemble and discusses strategies for sampling this ensemble using Monte Carlo (MC) simulations. The computation of interface free energies is demonstrated using a simple first-neighbor Ising model in \sect{sect:interfaces_ising}, followed by an analysis of convergence and scaling properties in \sect{sect:convergence}. Finally, in \sect{sect:interfaces_fecr} the VC-SGC MC approach is applied to extract interface free energies in Fe--Cr alloys as a function of temperature and orientation while taking into account both configurational and vibrational degrees of freedom. The paper is concluded in \sect{sect:conclusions}.

\section{Ensembles for multi-component systems}

We start this section with a review of the canonical and semi-grandcanonical ensembles before deriving the variance-constrained semi-grandcanonical (VC-SGC) ensemble. We conclude by discussing practical aspects pertaining to sampling the VC-SGC ensemble using Monte Carlo simulations.

\subsection{Canonical and semi-grandcanonical ensembles}

Let us start by a brief discussion of multiphase equilibria in multicomponent systems. For the sake of simplicity and clarity, we restrict ourselves to an immiscible binary alloy. The generalization to systems with an arbitrary number of species is straightforward. Consider a system of $N$ particles confined in a box of volume $V$ where each particle carries a spin of value 0 or 1. An arbitrary configuration is specified by $(\vec{x},\vec{\sigma})$, where $\vec{x}$ is a $3N$-dimensional vector describing the position of every particle in the box, and $\vec{\sigma}$ is an $N$-dimensional spin vector. The number of spin 1 particles is $n=\sum_{i=1}^N \sigma_i$, and their concentration is $c=n/N$. We denote the energy of a configuration by $\hat{U}(\vec{x},\vec{\sigma})$. The canonical partition function for this system at temperature $T$ reads
\begin{align}
  \mathcal{Z}_{\C}\left(c,\mathcal{N}\right) &
  = \frac{1}{n!(N-n)!} \int \exp\left[
    -\beta \hat{U}\left(\vec{x},\vec{\sigma}\right)\right] d\vec{x},
  \label{eq:Zc0}
\end{align}
where $\beta=1/k_BT$ and $\mathcal{N}=\{N,V,T\}$ is the set of independent thermodynamic variables. Using the identity
\begin{align}
  \frac{1}{n!(N-n)!} 
  &= \frac{1}{N!}\sum_{\sigma_1\cdots\sigma_N}\delta\left(\sum_{i=1}^N\sigma_i-n\right), 
 \label{eq:id}
\end{align}
we can rewrite the partition function in terms of an effective potential where the compositional degrees of freedom have been integrated out:
\begin{align}
  \mathcal{Z}_{\C}\left(c,\mathcal{N}\right)
  &= \frac{1}{N!} \int \exp\left[
    -\beta F_{\C}\left(\vec{x};c,\mathcal{N}\right)\right] d\vec{x}.
\label{eq:Zc}
\\
  F_{\C}\left(\vec{x};c,\mathcal{N}\right) &= -k_BT \ln 
  \sum_{\sigma_1\cdots
    \sigma_N}\delta\left(\sum_{i=1}^N\sigma_i-n\right)\nonumber
  \label{eq:fc}
  \\
  & \quad \times \exp\left[-\beta \hat{U}\left(\vec{x},\vec{\sigma}\right)\right].
\end{align}
$F_{\C}\left(\vec{x};c,\mathcal{N}\right)$ is the effective potential energy landscape for the structural degrees of freedom of the multicomponent system at temperature $T$. In this way, we separate the compositional and topological degrees of freedom. In the following, we focus our discussion on the contribution of the compositional degrees of freedom to the free energy.

We thus consider a frozen lattice of particles residing at $\vec{x}$. For brevity we drop all references to $\vec{x}$ and denote the potential energy of a particular spin configuration as $\hat{U}(\vec{\sigma})$. The SGC ensemble represents a set of configurations that sample the spin degrees of freedom according to the Boltzmann distribution while also allowing the total concentration to vary. To tune the average concentration an external chemical potential is applied to the system, which corresponds to modifying the potential energy function as follows 
\begin{align}
  \hat{U}_{\SGC}(\vec{\sigma}; \dmu)
  &= \hat{U}(\vec{\sigma}) + \dmu N\hat{c}(\vec{\sigma})
  \label{eq:u_sgc}
  \\
  \hat{c}(\vec{\sigma}) &= \sum_{i=1}^N\sigma_i/N,
\end{align}
where $\dmu$ is the relative chemical potential that controls the average concentration and the free energy is given by
\begin{align}
  F_{\SGC}\left(\dmu,\mathcal{N}\right) = -k_BT \ln 
  \sum_{\sigma_1\cdots \sigma_N}
  \exp\left[-\beta \hat{U}_{\SGC}\left(\vec{\sigma};\dmu\right)\right].
  \label{eq:fsgc}
\end{align}
Inserting \Eq{eq:u_sgc} into the above equation, the SGC partition function defined as $Z_{\SGC}=\exp\left(-F_{\SGC}/k_BT\right)$ can be expressed in terms of the canonical free energy as 
\begin{align}
  \mathcal{Z}_{\SGC}(\dmu,\mathcal{N}) &=
  \int_0^1 \exp\left[
    -\beta\left(F_{\C}(c,\mathcal{N})+\dmu Nc\right)\right] dc.
\label{eq:zsgc}
\end{align}
The integrand in the above equation can be used to define a concentration distribution function that is peaked around the average concentration $\left<\hat{c}\right>_{\SGC}$. The condition of zero derivative at $\left<\hat{c}\right>_{\SGC}$ yields the well-known thermodynamic relation 
\begin{align}
  \dmu &=
  -\frac{1}{N}\frac{\partial F_{\C}}{\partial c}
  \left(\left<\hat{c}\right>_{\SGC},\mathcal{N}\right).
  \label{eq:mu-f(c)}
\end{align}

\subsection{The variance constrained semi-grandcanonical ensemble}
 
Since for immiscible systems, one value of $\dmu$ maps to several compositions inside the miscibility gap stable multiphase coexistence cannot be established in the SGC ensemble. To overcome this restriction we modify the SGC ensemble in such a way as to control concentration fluctuations inside the miscibility gap. This is most easily accomplished by adding a constraint that fixes the ensemble-averaged squared concentration $\left<\hat{c}^2\right>$. We call this the variance-constrained semi-grand canonical (VC-SGC) ensemble.

This approach is analogous to umbrella sampling when calculating free energy barriers for reactions in molecular systems. \cite{TorVal77, KasThi05} In this method, given a particular reaction coordinate $\xi$, a harmonic external potential $u^b\left(\xi;\overline{\xi}\right) = K/2 \left(\xi-\overline{\xi}\right)^2$ is applied to bias the system toward positions $\overline{\xi}$ while at the same time the fluctuations are restrained by the force constant $K$. In the same spirit we can write the potential energy function of the VC-SGC ensemble as
\begin{align}
  \hat{U}_{\VCSGC}(\vec{\sigma}; \phi,\kappa) &=
  \hat{U}(\vec{\sigma})
  + \kappa \left( N\hat{c}(\vec{\sigma})
  + \phi/2\kappa \right)^2,
  \label{eq:u_vcsgc}
\end{align}
where $\phi$ and $\kappa$ are now thermodynamic variables that control the average concentration as well as its fluctuations. The harmonic external potential is parametrized such that $\hat{U}_{\VCSGC}\rightarrow \hat{U}_{\SGC}$ except for a constant shift whenever $\kappa\rightarrow 0$ and $\phi\rightarrow\dmu$. $\kappa$ can be given a physical intrepretation as the generalized force that controls concentration fluctuations. This can be realized by putting the system in contact with a finite reservoir. \cite{Het87, JohPlaViv03} Hence the VC-SGC ensemble, like the closely related extended Gaussian ensemble, is only applicable to finite systems. As all fluctuations vanish in the thermodynamic limit, so does $\kappa$. It should thus diminish as the system size grows. This poses no difficulty in our study of interfacial free energies since they also tend to zero in the thermodynamic limit. 

Analogously to the SGC ensemble [see Eqs.~\eq{eq:fsgc} and \eq{eq:zsgc}] we can express the VC-SGC partition function in terms of the canonical free energy as 
\begin{align}
  \mathcal{Z}_{\VCSGC}(\phi,\kappa,\mathcal{N})
  &= \int_0^1 \exp\left[
    -\beta\left(F_{\C}(c,\mathcal{N})+\Delta U_{\VCSGC}^b\right)\right] dc,
     \label{eq:zvcsgc} \\
  \Delta U_{\VCSGC}^b
  &= \kappa\left(Nc+\phi/2\kappa\right)^2,
\end{align}
where $\Delta U_{\VCSGC}^b$ corresponds to the harmonic bias potential of the umbrella sampling approach. The integrand in \Eq{eq:zvcsgc} describes the probability distribution of the global concentration in the VC-SGC ensemble. It is a peaked function around the average concentration $\left<\hat{c}\right>_{\VCSGC}$. The condition of zero derivative at $\left<\hat{c}\right>_{\VCSGC}$ yields the following relation between $\phi$ and $\kappa$ and the canonical free energy for the VC-SGC ensemble
\begin{align}
  \phi + 2N\kappa\left<\hat{c}\right>_{\VCSGC} &=
  -N^{-1}\partial F_{\C}/\partial c
  \left(\left<\hat{c}\right>_{\VCSGC},\mathcal{N}\right).
  \label{eq:mu-v(c)}
\end{align}

It is important to note that the right-hand side of \Eq{eq:mu-v(c)} is the concentration derivative of the canonical free energy at $\left<\hat{c}\right>_{\VCSGC}$, which is a simulation observable. Hence no more unbiasing is necessary when employing the above technique for free energy integration. However, for very small system sizes where the condition of zero derivative at $\left<\hat{c}\right>_{\VCSGC}$ may not hold, the simple procedure outlined above must be modified. The standard approach in such cases is to extract the probability distribution of the system as a function of concentration via histogram methods \cite{FerSwe88, FerSwe89, Rou95} and sampling bias must be corrected for as in the usual implementations of umbrella sampling.\cite{TorVal77, KasThi05, DicLegLel10}.

Notwithstanding, the key advantage of the VC-SGC ensemble is that while the chemical driving force (right-hand side of \Eq{eq:mu-v(c)}) is an ensemble observable, the constraint on concentration fluctuations simultaneously allows multiphase equilibria to be stabilized. This is simpler and more efficient than the state-of-the-art methods for calculating concentration-derivatives of the free energy within the canonical ensemble, which measure the work required to perform a gradual/instantaneous transmutation of one species into the other while keeping the particle number fix. While quite clever path-sampling algorithms with optimized estimators for calculating this work have been devised, \cite{AdjAthRod11} for the particular application discussed in this paper, the variance-constrained ensemble presents a simpler and more efficient route, since the free energy derivative can be obtained as an observable of the equilibrium ensemble, and no calculation of external work is required. Furthermore as discussed in detail in a previous publication, \cite{SadErhStu12} the VC-SGC MC method allows for faster convergence to equilibrium in multiphase regions of the phase diagram than the canonical ensemble. Finally the VC-SGC ensemble is also very well suited for efficient parallel MC algorithms \cite{SadErhStu12} that enable simulations of very large systems containing millions of particles.

\subsection{Efficient sampling of the VC-SGC ensemble}
\label{sect:practical_aspects}

When carrying out MC simulations in the SGC ensemble according to \Eq{eq:mu-f(c)}, the parameter $\Delta\mu$ is directly proportional to the free energy derivative $\partial F/\partial c$, which renders choosing suitable parameters straightforward. We will now demonstrate that choosing the relevant range of values for VC-SGC simulations is just as simple as in the case of the SGC ensemble.

In the previous section the VC-SGC ensemble was derived in terms of the parameters $\phi$ and $\kappa$, which led to expressions that very much resembled the SGC ensemble. In practice it turns out that it is more convenient to use the substitutions $\kappa=\bar{\kappa}/N$ and $\phi=\bar{\kappa}\bar{\phi}$. The VC-SGC potential defined above in \Eq{eq:u_vcsgc} then reads
\begin{align}
  \hat{U}_{\VCSGC}(\vec{\sigma}; \bar{\phi},\bar{\kappa})
  &=
  \hat{U}(\vec{\sigma}) + \bar{\kappa} N \left( c + \bar{\phi}/2 \right)
\end{align}
and the associated expression for the first derivative of the free energy [compare \Eq{eq:mu-v(c)}] is
\begin{align}
  \bar{\kappa} \left(\bar{\phi} + 2 \left<\hat{c}\right>_{\VCSGC} \right)
  &=
  -N^{-1}\partial F_{\C}/\partial c
  \left(\left<\hat{c}\right>_{\VCSGC},\mathcal{N}\right).
  \label{eq:dF}
\end{align}
The above transformations effectively decouple the average condtraint from the variance constraint parameter, which simplifies parameter selection for VC-SGC MC simulations as will be seen in the following.

It is instructive to first consider a symmetric free energy profile, in which case $\partial F/\partial c=0$ at the solubility limits where $F$ is minimal as well as around 50\%\ where $F$ is maximal. From \Eq{eq:dF} it follows that for non-zero $\bar{\kappa}$ one can thus install a concentration of 50\%\ by choosing $\bar{\phi}=-1$. Furthermore one obtains $\left<\hat{c}\right>_{\VCSGC}\rightarrow 1$ for $\bar\phi\approx -2$ and $\left<\hat{c}\right>_{\VCSGC}\rightarrow 0$ for $\bar\phi \approx 0$. These observations suggest a simple protocal for an efficient sampling of the full concentration range: First choose a constant value for $\bar\kappa$ and then vary $\bar\phi$ in the range $-2.2\lesssim\bar\phi\lesssim 0.2$. In our experience this approach is very transferable and we have used it successfully in exactly this fashion for various systems including for example the (very asymmetric) Fe--Cr system as described in \sect{sect:interfaces_fecr}.

In the following section it will be shown among over things that the results for the free energy derivative are insensitive to the variance constraint parameter $\bar\kappa$ over a wide range of values. This provides for a lot of freedom in choosing $hat\kappa$.

It is straightforward to formulate a Monte Carlo algorithm \cite{SadErhStu12} for sampling the VC-SGC ensemble, where transmutation trial moves comprise
\begin{enumerate}
\item selecting a particle at random
\item flipping its spin (type) and
\item computing the energy change $\Delta E$ and the concentration change $\Delta c$
\end{enumerate}
These trial moves are then accepted with probability
\begin{align}
  \mathcal{A}_\VCSGC = \min \left\{ 1,
  \exp\left[ - \beta \left(
    \Delta E + \bar\kappa N c \left( \bar\phi + \Delta c + 2 c \right)
    \right) \right]
  \right\},
\end{align}
which satisfies detailed balance.

\section{Basic features of the VC-SGC MC technique}
\label{sect:basic_features}

\subsection{Interface free energies from direct thermodynamic integration}
\label{sect:interfaces_ising}

\begin{figure}
  \centering
\includegraphics[width=\columnwidth]{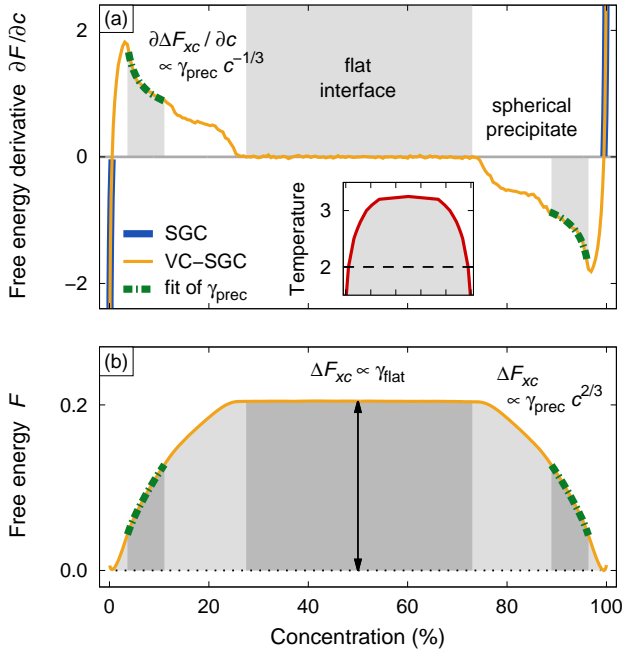}
  \caption{
    (Color online)
    (a) Derivative of free energy with respect to concentration at $T=1.5\,k_B$ for system size $8\times8\times12$. The gray areas indicate composition regions, in which spherical precitates and flat interfaces are formed. The bold green lines indicate fits to \Eq{eq:gamma_sphere}. The inset shows the phase diagram for the Ising model described in the text, where temperature is measured in units of $k_B$.
    (b) Free energy profile obtained by integration of data in (a).
  }
  \label{fig:phasediagram}
  \label{fig:freen}
\end{figure}

In this section we illustrate the utility of the VC-SGC ensemble by studying phase segregation in a simple Ising model. We will calculate excess free energies of interfaces between two coexisting phases by direct thermodynamic integration from the single-phase region of the phase diagram into the miscibility gap.  The Ising model is defined on a body-centered cubic (BCC) lattice with its Hamiltonian parametrized as follows 
\begin{align}
  \mathcal{H} = -\frac{1}{2} \sum_{ij} S_i S_j,
\end{align}
where the summation runs over first-nearest neighbor pairs and $S_i \in \left\{-1,1\right\}$. The phase diagram of this system is shown in the inset of \fig{fig:phasediagram}(a).

Our first goal in this section is to show that the VC-SGC ensemble, unlike the SGC ensemble, allows for calculation of free energy derivatives in both single and multiphase regions of the phase diagram. To this end, simulations were carried out using cells with basis vectors oriented along $[100]$, $[010]$, and $[001]$. In the VC-SGC simulations the constraint $\bar\phi$ (controlling average concentration) was varied from $-2.05$ to 0.05 in steps of 0.01 at a constant variance constraint of $\bar\kappa = 100$. In case of SGC simulations the chemical potential difference $\dmu$ was changed between $-3$ and $+3$. At each value of $\bar\phi$ (or $\dmu$) the system was equilibrated for $5\times10^3$ MC sweeps, after which statistics were gathered for $15\times10^3$ MC sweeps.

\begin{figure}
  \centering
\includegraphics[width=0.9\columnwidth]{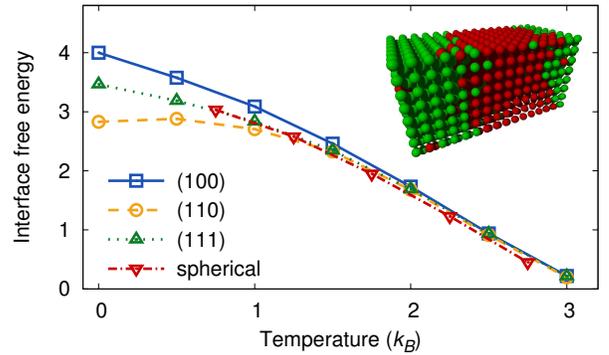}
  \caption{
    (Color online)
    Interface free energies as a function of temperature and orientation for the Ising model described in the text.
    The inset shows a snapshot of a $\left\{100\right\}$ interface from a VC-SGC MC simulation close to 50\%\ composition (picture rendered with \textsc{ovito}, Ref.~\onlinecite{Stu10}).
  }
  \label{fig:interfaces}
  \label{fig:interface_orientations}
\end{figure}

The thus obtained relations between free energy derivative $\partial F/\partial c$ and concentration $c$ are shown in  \fig{fig:phasediagram}(a) for a temperature of $1.5\,k_B$. The gap in the blue line in \fig{fig:phasediagram}(a) corresponds to the miscibility gap. This reflects the inability of the SGC ensemble in stabilizing multiphase equilibria. The two concentrations delineating the boundary of the miscibility gap, i.e. 1\%\ and 99\%\ in \fig{fig:phasediagram}, are commonly referred to as binodals.

Using the VC-SGC MC method $F'(c)$ can be calculated for {\em all} concentrations as shown by the yellow line in \fig{fig:freen}(a). Three regimes can be distinguished: At concentrations between the pure phases and the binodals (regime I), single-phase equilibria are obtained and the SGC-MC results are reproduced. In this regime, $F'(c)$ has a positive slope. This behavior is preserved for compositions between binodal and critical nucleation (extrema in $\partial F/\partial c$ in \fig{fig:freen}(a)), which only support formation of subcritical minority-phase clusters (regime II). In this regime there is a finite thermodynamic driving force for phase segregation but due to small sizes of the minority-phase clusters the interfacial free energy dominates and leads to their overall formation free energies being positive. Finally at concentrations beyond critical nucleation (regime III), supercritical clusters are formed and $F'(c)$ is a monotonically decreasing function.  In this regime the excess free energy increases due to the increase in the interface area of the cluster. Note that the van-der-Waals loop obtained in \fig{fig:freen}(a) is the equilibrium solution under the constraint of finite system size. In particular, the concentration range spanned by regime II diminishes as the simulation cell increases, and eventually vanishes in the thermodynamic limit.

Studing the shapes of the supercritical clusters in regime III, we find several subregimes corresponding to different types of precipitates. For concentrations immediately beyond critical nucleation, minority-phase clusters are compact. The excess free energy (proportional to the interface area) as a function of concentration can thus be written as
\begin{align}
  \Delta F_{xc}(c)
  &= \xi \left[V_\text{cell}(c-c_0)\right]^{2/3} \gamma_\text{prec}.
  \label{eq:gamma_sphere}
\end{align}
Here, $\xi$ is a shape factor, which equals $(36\pi)^{\frac{1}{3}}$ for spherical precipitates, and $V_\text{cell}$ denotes the volume of the simulation cell. The bold dashed green lines in \fig{fig:phasediagram} indicate fits to \Eq{eq:gamma_sphere} and its concentration derivative. The effective interface free energies calculated in this way are shown as a function of temperature in \fig{fig:interfaces}.

\begin{figure}
  \centering
\includegraphics[width=\columnwidth]{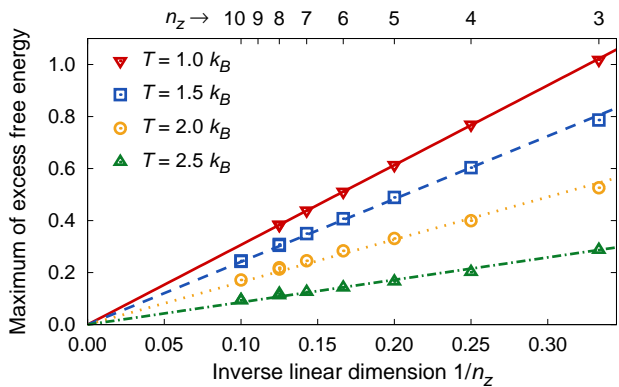}
  \caption{
    (Color online)
    Scaling of maximum of excess free energy [compare \fig{fig:phasediagram}(b)] with inverse linear dimension for the Ising model described in the text. Simulation cell vectors are oriented along $\left<100\right>$ directions inducing the formation of $\left\{100\right\}$ interfaces.
  }
  \label{fig:Fmax_scaling}
\end{figure}

At still higher concentrations, the precipitates become so large that spherical clusters can no longer be contained inside the finite-sized simulation cell and a transition to a cylindrical shape induced by periodic boundary conditions is observed. While the free energy profile recorded in this region is of limited physical interest {\it per se}, it provides the continuous derivative needed for integration of the full excess free energy. 

Finally in the concentration range around 50\%\ in \fig{fig:freen}(a), the excess free energy assumes a constant value due to formation of superlattices consisting of alternating slabs of the two phases, which follow the periodicity of the simulation cell. The excess free energy at these concentrations corresponds to the free energy cost associated with two flat interfaces with the free energy density given by 
\begin{align}
  \gamma_\text{flat} = \Delta F_{xc} / A,
\end{align}
where $A$ is the cross-sectional area of the computational cell. By changing the geometry of the simulation cell it is possible to change $A$ and thereby interface orientation. To see this remember that during the course of a simulation the system will establish the interface configuration that minimizes $\Delta F_{xc}$ not $\gamma_{\text{flat}}$. Since the cross section $A$ for different orientations can be changed by varying shape and size of the simulation cell, it is possible to stabilize different interface orientations as exemplified in \fig{fig:interface_orientations}. To obtain for example the free energy density of $\left\{111\right\}$ interfaces, we used an orthorhombic simulation cell with lattice vectors oriented along $\left[1\bar{1}0\right]$, $\left[\bar{1}\bar{1}2\right]$, and $\left[111\right]$ and $2\times3\times8$ unit cells. In this configuration the excess free energy $\Delta F_{c}$ for $\left\{100\right\}$ and $\left\{110\right\}$ interfaces is larger than for $\left\{111\right\}$ interfaces. Similarly shortening the dimensions perpendicular to $\left[1\bar{1}0\right]$ and extending the cell parallel to this direction stabilizes a $\left\{110\right\}$ interface. In this fashion it is possible to not only extract the temperature but also the orientation dependence of the interface free energy, which leads to the data displayed in \fig{fig:interfaces}.

\begin{figure}
  \centering
\includegraphics[width=\columnwidth]{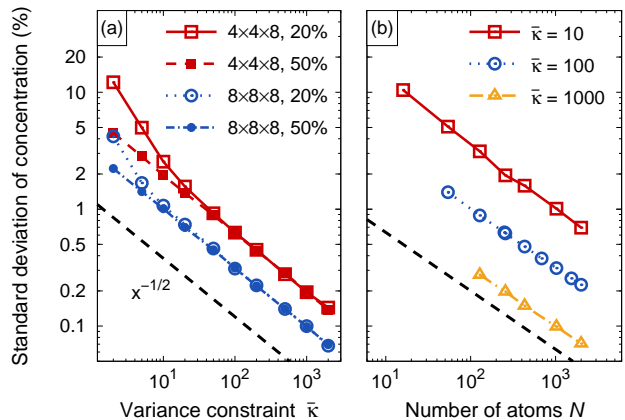}
  \caption{
    (Color online)
    Standard deviation of concentration at a temperature of $T=2\,k_B$as a function of (a) variance constraint parameter and (b) system size. Data in (b) measured at 50\%\ composition. The black dashed line indicates scaling with $1/\sqrt{x}$.
  }
  \label{fig:dc_kappa_nat}
\end{figure}

To verify the reliability of the thus computed interfacial free energies using the VC-SGC MC technique, we have also studied the dependence of the excess free energies of the two-phase equilibria inside the miscibility gap on the size of the simulation cell. Since the excess free energy is dominated by the interfacial free energy it should scale inversely with the longest linear dimension of the simulation cell. This behavior is indeed observed as shown in \fig{fig:Fmax_scaling}.

\subsection{Convergence and parameter sensitivity}
\label{sect:convergence}

In this section we study the convergence properties of VC-SGC MC simulations and the sensitivity of the calculated free energies to the choice of the variance constraint parameter. Figure~\ref{fig:dc_kappa_nat} shows the dependence of the standard deviation of concentration on (a) variance constraint parameter and (b) system size. In both cases the standard deviation scales inversely with the square root of the respective parameter. To understand this behavior consider again \Eq{eq:zvcsgc}. The standard deviation of concentration in the VC-SGC ensemble can be expressed as the second moment of its probability distribution as follows
\begin{align}
  s_V^2 =   \frac{1}{\mathcal{Z}_{\VCSGC}}
  \int_0^1 c^2 \exp\left[
    -\beta\left(F_{\C}(c,\mathcal{N})+\Delta U_{\VCSGC}^b\right)\right] dc
  - \left<\hat{c}\right>_V^2.
  \label{eq:zc}
\end{align}
For large enough system sizes, the ensemble probability distribution becomes a normal distribution $\sqrt{A/\pi}\exp\left(-A(c-B)^2\right)$, with its mean $B=\left<\hat{c}\right>_V$, and its width $A$ inversely proportional to the standard deviation: $s_V^2=1/2A$. Now the Gaussian width $A$ is related to the second derivative of the distribution taken at the average concentration. We thus obtain the following relation for $s_V^2$
\begin{align}
  \frac{1}{\beta s_V^2} = \frac{\partial^2 F_{\C}\left(\left<\hat{c}\right>_V,\mathcal{N}\right)}{\partial c^2}+2\bar\kappa N
  \label{eq:sV_dF}
\end{align}
This is a very important relation that clearly predicts the dependence of the standard deviation of concentration in the VC-SGC ensemble on the variance constraint parameter $\bar\kappa$ as well as on the system size $N$ as observed in Figs.~\ref{fig:dc_kappa_nat}(a) and \ref{fig:dc_kappa_nat}(b). It also provides a clear picture of why and under which conditions the VC-SGC ensemble can stabilize multiphase equilibria. Equation~\eq{eq:sV_dF} illustrates that whenever $\bar\kappa\rightarrow 0$ as in the SGC ensemble, $s_V$ diverges for values of $\left<\hat{c}\right>_V$, for which the second derivative of the free energy function $F_{\C}\left(c,\mathcal{N}\right)$ becomes negative. For these compositions, stable single-phase equilibria cannot be achieved. Introducing a non-zero variance constraint $\bar\kappa$ leads to finite $s_V$. However, in order for $\bar\kappa$ to stabilize compositions inside the miscibility, it has to be chosen large enough to make the right-hand side of \Eq{eq:sV_dF} positive.  Hence for practical calculations, the best choice for $\bar\kappa$ is a value slightly larger than the maximum of $\frac{1}{2N}\left|F''_{\C}\left(c,\mathcal{N}\right)\right|$ inside the miscibility gap. According to \fig{fig:dc_kappa_nat}(a) this threshold is about $\bar{\kappa}=20$ for a system with $4\times4\times8$ unit cells and close to 10 for a size of $8\times8\times8$ unit cells.

For $\kappa\rightarrow0$ the VC-SGC ensemble approaches the SGC ensemble, for which the standard deviation should diverge for concentrations inside the miscibility gap, a behavior that is in fact clearly visible in \fig{fig:dc_kappa_nat}. The figure also explicitly demonstrates that in the limits $\bar\kappa\rightarrow\infty$ and/or $N\rightarrow\infty$ concentration fluctuations go to zero and thus approach the canonical ensemble.

\begin{figure}
  \centering
\includegraphics[width=\columnwidth]{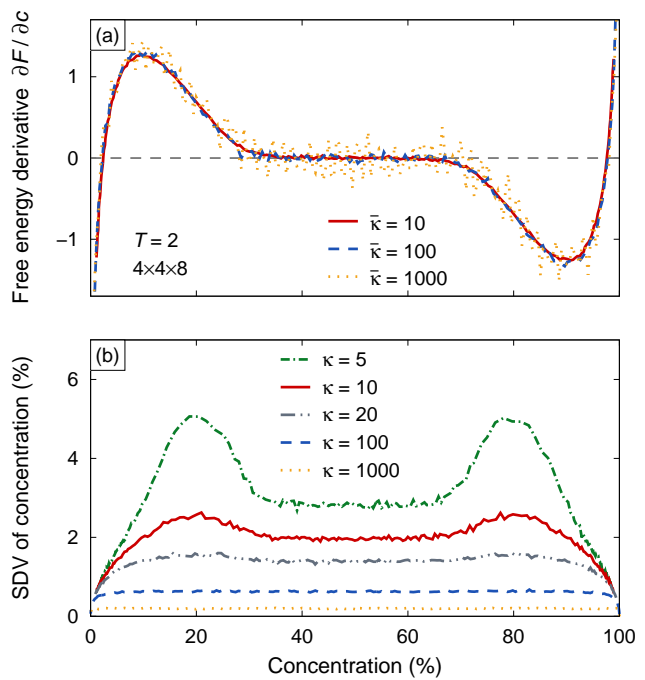}
  \caption{
    (Color online)
    Effect of variance constraint parameter $\bar\kappa$ on (a) free energy derivative and (b) standard deviation of concentration at $T=2\,k_B$ for system size $4\times4\times8$.
  }
  \label{fig:dF_sdv}
\end{figure}

\begin{figure}
  \centering
\includegraphics[width=\columnwidth]{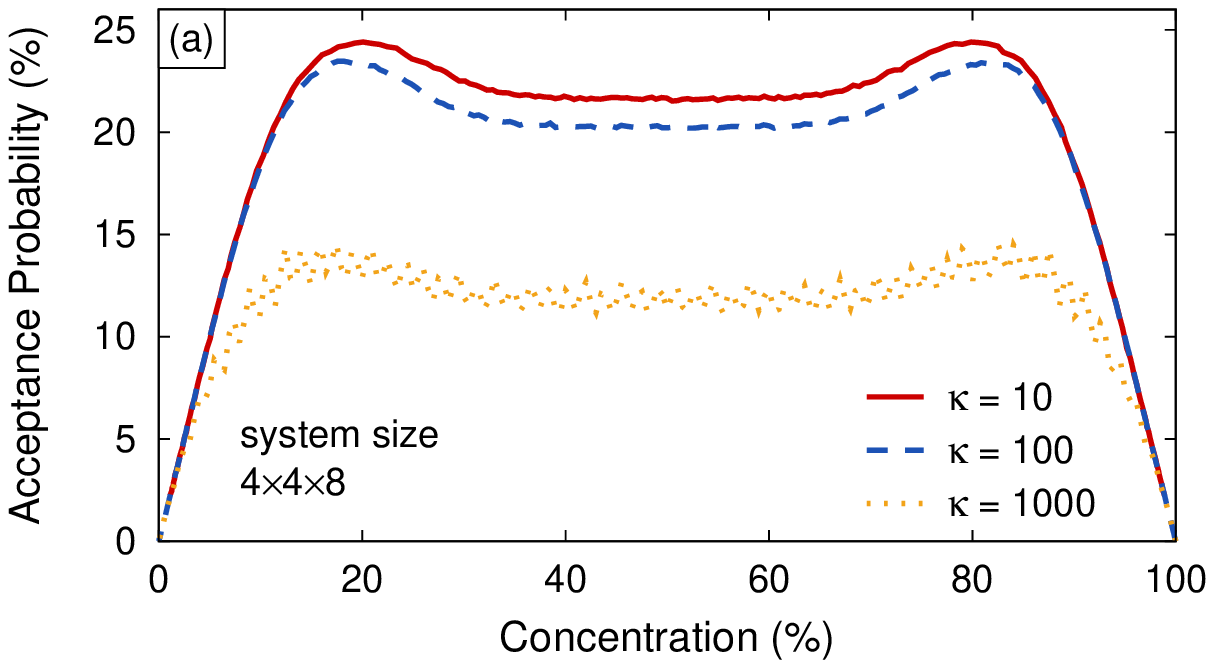}
\includegraphics[width=\columnwidth]{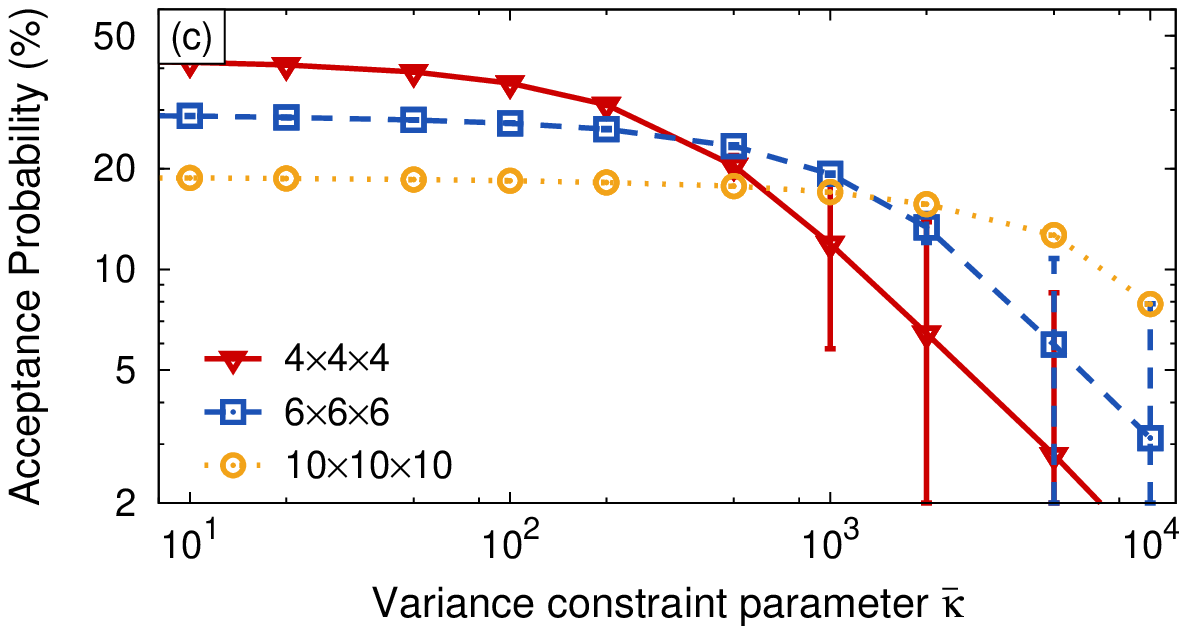}
  \caption{
    (Color online)
    Effect of variance constraint parameter $\bar\kappa$ on (a) free energy derivative and (b) standard deviation of concentration at $T=2\,k_B$ for system size $4\times4\times8$.
  }
  \label{fig:accprob}
\end{figure}

Figure~\ref{fig:dF_sdv}(a) illustrates the important fact that the free energy derivative obtained via \Eq{eq:mu-v(c)} is indeed {\em unaffected} by the strength of the variance constraint over a wide range of parameter values. The lower limit of the acceptable parameter range is set by the divergence of the standard deviation that was described in the previous paragraph and that is shown as a function of concentration in \fig{fig:dF_sdv}(b). The upper limit, however, is due to decreasing acceptance probability of the MC trial moves, which as illustrated in \fig{fig:accprob} diminishes only slightly with $\bar\kappa$ over several orders of magnitude before dropping more rapidly. As a result for large values of $\bar\kappa$ one can no longer gather sufficient statistics to compute meanigful thermodynamic averages. The range within which $\bar\kappa$ can be selected nonetheless extends at least over two to three orders of magnitude depending on system size. Since large $\bar\kappa$ values lead to a reduction of the acceptance probability, it is, however, recommended to choose $\bar\kappa$ as small as possible while maintaining an acceptable standard deviation. Figure~\ref{fig:dc_kappa_nat} suggests that a reasonable value is about 1\%.

\section{Interface free energies in the Fe--Cr alloy system}
\label{sect:interfaces_fecr}

\begin{figure}
  \centering
\includegraphics[width=\columnwidth]{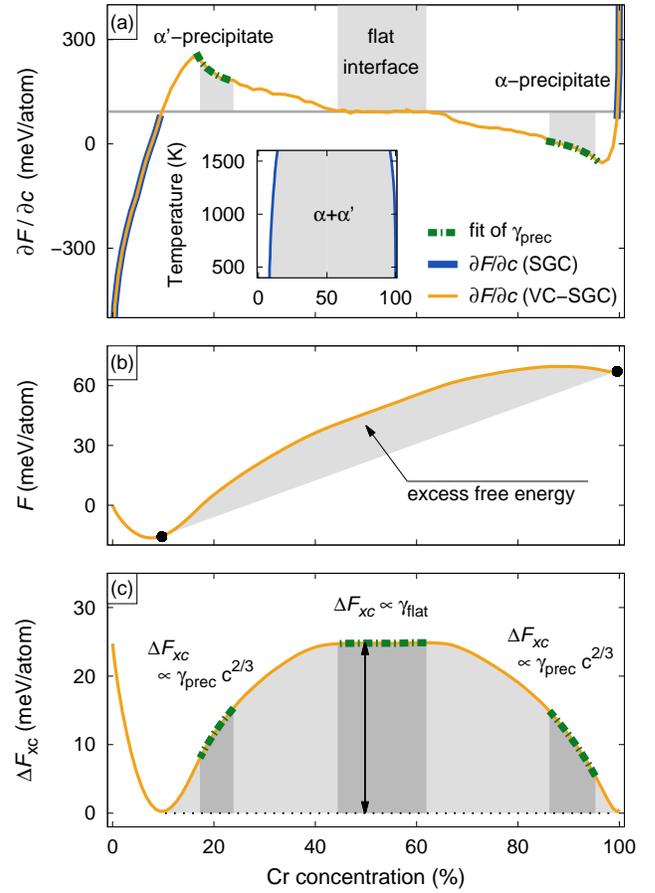}
  \caption{
    (a) Derivative of free energy with respect to concentration obtained from MC simulations in the SGC and VC-SGC ensembles at \text{1000\,\K}. The inset shows the calculated phase diagram, which is in excellent agreement with the published phase diagram of the potential used in the present work. \cite{BonErhCar09}
    (b) Free energy profile obtained by integration of the data in (a) and (c) the corresponding excess free energy.
    The central dark gray area indicates the region within which flat interfaces are obtained while the dark gray areas to the left and right show the regions in which compact precipitates are obtained.
  }
  \label{fig:freen_FeCr}
\end{figure}

Up until now we have considered a simple Ising model. It can be considered as the most simple representative of alloy cluster expansions, \cite{SanDucGra84, AstOzoWoo01} which are widely used to study complex multi-component alloys. The VC-SGC MC technique is, however, equally applicable to empirical potential models and then enables computation of excess free energies taking into account not only configurational but also structural and vibrational degress of freedom. \footnote{
  In general, the two phases that are separated by the interface can exhibit size mismatch leading to an additional strain contribution to the interface (free) energy.\cite{Mis04, MisAstLi10} This contribution can be added to the present approach without further complications. In the case of the Fe--Cr system the size mismatch is miniscule and the strain contribution is neglible.
}
To illustrate this point, we now consider the calculation of excess free energies in BCC the Fe--Cr binary alloy system as described by the concentration-dependent embedded atom method. \cite{CarCroCar05, StuSadErh09} The latter reproduces the phase diagram below the Curie temperature featuring both substantial solubility of Cr in Fe ($\alpha$-phase) and a pronounced miscibility gap as shown in the inset of \fig{fig:freen_FeCr}(a).

Simulations were carried out using orthorhombic cells containing between 384 and 480 atoms depending on sample orientation. The variance constraint parameter was set to $\bar\kappa$=100\,eV/atom and $\bar\phi$ was varied between $-2.1$ and 0.1 (see \sect{sect:practical_aspects}). Structural relaxations and thermal vibrations were sampled using displacement MC trial moves, while volume trial moves were employed to sample zero pressure conditions. The system was evolved for 6,000 MC steps at each value of $\bar\phi$. Statistics were gathered over the last 5,000 steps.

First we mapped out $\partial F/\partial c$ using SGC MC simulations as shown for a temperature of 1000\,K by the bold blue line in \fig{fig:freen_FeCr}(a). The gap in the blue line corresponds to the miscibility gap, which again reflects the inability of the SGC ensemble to stabilize multiphase equilibria. Using the VC-SGC MC method we then calculated $\partial F/\partial c$ over the {\em entire} concentration range as shown by the yellow line in Fig.~\ref{fig:freen_FeCr}(a). The free energy profile obtained in this fashion again coincides with the SGC results in the single phase regions and inside the miscibility gap shows the same features that were discussed in \sect{sect:interfaces_ising} for the Ising model.

\begin{figure}
  \centering
\includegraphics[width=\columnwidth]{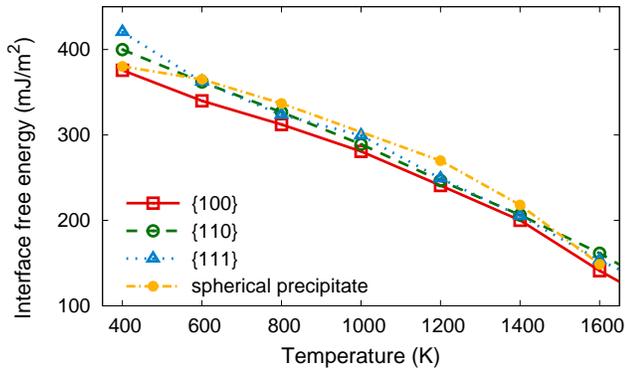}
  \caption{
    Interface free energies for Fe--Cr model alloy for different interface orientations as well as spherical precipitates.
  }
  \label{fig:gsurf}
\end{figure}

Once again it is possible to extract free energy densities corresponding to spherical and flat interfaces from the excess free energy curves, as shown in \fig{fig:freen_FeCr}(c). The results of this analysis are summarized in \fig{fig:gsurf}. Note that in this case the computation of zero-K interface energies is significantly more complicated than in the case of the Ising model. The asymmetric phase diagram features a large solubility of Cr in $\alpha$-Fe that remains finite as temperature goes to zero and is accompanied by short-range ordering. \cite{ErhCarCar08} We have therefore not attempted to construct a zero-K $\alpha/\alpha'$ interface.

Two features in \fig{fig:gsurf} are particularly noteworthy in comparison with \fig{fig:interfaces}. The ordering of the interface orientations is different with $\left\{100\right\}$ interfaces now being lowest in energy, and the anisotropy is much smaller.

\begin{figure}
  \centering
\includegraphics[width=\columnwidth]{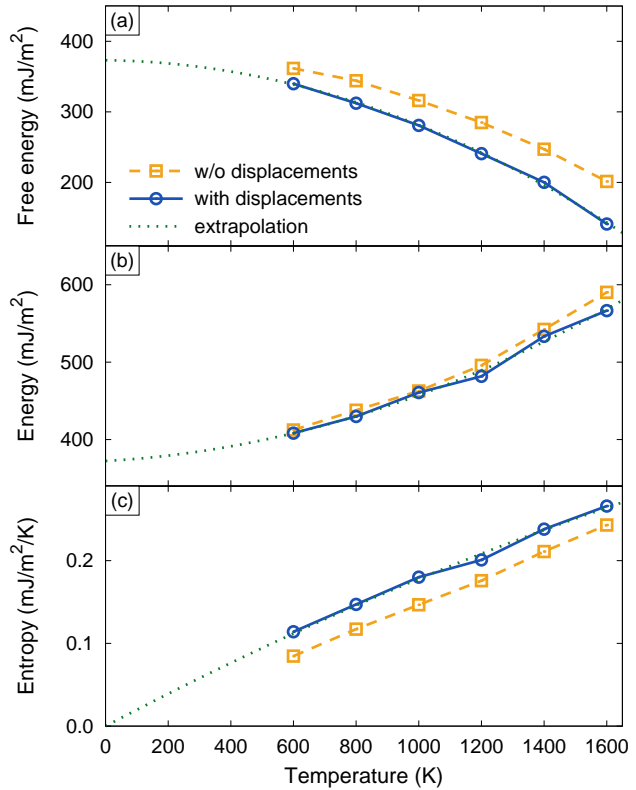}
  \caption{
    (a) Free energy, (b) internal energy, and (c) entropy of $\left\{100\right\}$ interface in Fe--Cr model alloy. The dashed lines represent fits to second degree polynomials.
  }
  \label{fig:surf_comp}
\end{figure}

It is furthermore instructive to separate the different contributions to the interface free energy. First, by repeating the calculations for $\left\{100\right\}$ interfaces {\em without} displacement trials moves\footnote{
  Volume change trial moves were still included in order to allow the system to relax lattice strain.
}
it is possible to separate vibrational and configurational degrees of freedom. The result of this analysis is shown in \fig{fig:surf_comp}(a), which illustrates the magnitude of the vibrational contribution to the interface free energy. Since the simulations also provide average internal energies one can furthermore separate explicitly the interface internal energies and entropies. As shown in \fig{fig:surf_comp}(b) and (c) the internal energies are hardly affected if vibrations and atomic relaxations (other than volume changes) are suppressed. Instead their contributions show up as a practically temperature independent contribution to the entropy. Figures~\ref{fig:surf_comp}(a) and (b) finally show that interface free and internal energies extrapolate to a value of about $372\pm2\,\text{mJ/m}^2$ at zero K.

\section{Conclusions}
\label{sect:conclusions}

In this paper we have introduced a free energy integration technique along compositional degrees of freedom that allows for efficient calculations of free energies of multiphase systems with various degrees of heterogeneity. The construction of the underlying VC-SGC ensemble represents a generalization of the extended Gaussian ensemble technique to multicomponent systems. \cite{Het87, JohPlaViv03}

To illustrate the basic features of the VC-SGC MC technique as well as aspects such as scaling and convergence, an extensive characterization of a first-nearest neighbor Ising model was carried out. In this way it was shown that the VC-SGC MC technique allows ({\it i}) stabilizing arbitrary concentrations inside (and outside) the miscibility gap and ({\it ii}) simultaneously computing derivatives of the free energy with respect to concentration. Efficient sampling of the entire concentration range is straightforward since in essence it merely requires setting the variance parameter $\bar\kappa$, which can be chosen rather deliberately.

Since VC-SGC MC simulations provide free energy derivatives as a continuous function of composition, thermodynamic integration can be carried out along the concentration axis, which allows determination of interface free energy densities. Specifically by monitoring the free energy derivative as precipitate size and shape evolve, it is possible to extract the excess free energies of compact precipitates. By pushing the system close to 50-50 phase composition while varying simulation cell shape and size, one can also derive the interface free energies of flat interfaces as a function of orientation.

To demonstrate the applicability of the VC-SGC approach in more general cases, we finally considered Fe--Cr alloys as described by an empirical potential scheme. The interface free energies for this system were determined taking into account not only the configurational but also vibrational and structural degrees of freedom. Our results indicate a strong temperature but weak orientation dependence of the interface free energies, suggesting that both $\alpha$ and $\alpha'$ precipitates should adopt a spherical shape at practically all relevant temperatures.

As described in Ref.~\onlinecite{SadErhStu12}, the VC-SGC MC ensemble leans itself to efficient parallelization, enabling simulations of precipitation in multi-million atom samples. \cite{ErhMarSad12}
It should finally be stressed that our approach is also directly applicable for obtaining interface free energies and nucleation barriers from cluster-expansions, \cite{SanDucGra84, AstOzoWoo01} which are widely applied to study complex multi-component alloys.

\begin{acknowledgments}
We would like to thank G. Gilmer at LLNL and M. Ath\`enes at CEA-Saclay for helpful discussions. Lawrence Livermore National Laboratory is operated by Lawrence Livermore National Sec\-urity, LLC, for the U.S. DOE-NNSA under Contract DE-AC52-07NA27344. P.E. has been partly funded by the Swedish Research Council. Com\-puter time allocations by NERSC at Lawrence Berkeley National Laboratory and by the Swedish National Infractstructure for Computing are gratefully acknowledged.
\end{acknowledgments}


\begin{thebibliography}{10}%
\makeatletter
\providecommand \@ifxundefined [1]{%
 \ifx #1\undefined \expandafter \@firstoftwo
 \else \expandafter \@secondoftwo
\fi
}%
\providecommand \@ifnum [1]{%
 \ifnum #1\expandafter \@firstoftwo
 \else \expandafter \@secondoftwo
\fi
}%
\providecommand \enquote [1]{``#1''}%
\providecommand \bibnamefont  [1]{#1}%
\providecommand \bibfnamefont [1]{#1}%
\providecommand \citenamefont [1]{#1}%
\providecommand\href[0]{\@sanitize\@href}%
\providecommand\@href[1]{\endgroup\@@startlink{#1}\endgroup\@@href}%
\providecommand\@@href[1]{#1\@@endlink}%
\providecommand \@sanitize [0]{\begingroup\catcode`\&12\catcode`\#12\relax}%
\@ifxundefined \pdfoutput {\@firstoftwo}{%
 \@ifnum{\z@=\pdfoutput}{\@firstoftwo}{\@secondoftwo}%
}{%
 \providecommand\@@startlink[1]{\leavevmode}%
 \providecommand\@@endlink[0]{}%
}{%
 \providecommand\@@startlink[1]{%
  \leavevmode
  \pdfstartlink
   attr{/Border[0 0 1 ]/H/I/C[0 1 1]}%
   user{/Subtype/Link/A<</Type/Action/S/URI/URI(#1)>>}%
  \relax
 }%
 \providecommand\@@endlink[0]{\pdfendlink}%
}%
\providecommand \url  [0]{\begingroup\@sanitize \@url }%
\providecommand \@url [1]{\endgroup\@href {#1}{\urlprefix}}%
\providecommand \urlprefix [0]{URL }%
\providecommand \Eprint[0]{\href }%
\@ifxundefined \urlstyle {%
  \providecommand \doi [1]{doi:\discretionary{}{}{}#1}%
}{%
  \providecommand \doi [0]{doi:\discretionary{}{}{}\begingroup
  \urlstyle{rm}\Url }%
}%
\providecommand \doibase [0]{http://dx.doi.org/}%
\providecommand \Doi[1]{\href{\doibase#1}}%
\providecommand \bibAnnote [3]{%
  \BibitemShut{#1}%
  \begin{quotation}\noindent
    \textsc{Key:}\ #2\\\textsc{Annotation:}\ #3%
  \end{quotation}%
}%
\providecommand \bibAnnoteFile [2]{%
  \IfFileExists{#2}{\bibAnnote {#1} {#2} {\input{#2}}}{}%
}%
\providecommand \typeout [0]{\immediate \write \m@ne }%
\providecommand \selectlanguage [0]{\@gobble}%
\providecommand \bibinfo [0]{\@secondoftwo}%
\providecommand \bibfield [0]{\@secondoftwo}%
\providecommand \translation [1]{[#1]}%
\providecommand \BibitemOpen[0]{}%
\providecommand \bibitemStop [0]{}%
\providecommand \bibitemNoStop [0]{.\EOS\space}%
\providecommand \EOS [0]{\spacefactor3000\relax}%
\providecommand \BibitemShut [1]{\csname bibitem#1\endcsname}%
\bibitem{BroLeFin08}%
  \BibitemOpen
  \bibfield{author}{%
  \bibinfo {author} {\bibfnamefont{Q.}~\bibnamefont{Bronchart}}, \bibinfo
  {author} {\bibfnamefont{Y.}~\bibnamefont{Le~Bouar}},\ and\ \bibinfo {author}
  {\bibfnamefont{A.}~\bibnamefont{Finel}},\ }%
  \bibfield{journal}{%
  \bibinfo {journal} {Phys. Rev. Lett.}\ }%
  \textbf{\bibinfo {volume} {100}},\ \bibinfo {pages} {015702} (\bibinfo {year}
  {2008})%
  \bibAnnoteFile{NoStop}{BroLeFin08}%
\bibitem{BroGil86c}%
  \BibitemOpen
  \bibfield{author}{%
  \bibinfo {author} {\bibfnamefont{J.~Q.}\ \bibnamefont{Broughton}}\ and\
  \bibinfo {author} {\bibfnamefont{G.~H.}\ \bibnamefont{Gilmer}},\ }%
  \bibfield{journal}{%
  \bibinfo {journal} {J. Chem. Phys.}\ }%
  \textbf{\bibinfo {volume} {84}},\ \bibinfo {pages} {5759} (\bibinfo {year}
  {1986})%
  \bibAnnoteFile{NoStop}{BroGil86c}%
\bibitem{DavLai03}%
  \BibitemOpen
  \bibfield{author}{%
  \bibinfo {author} {\bibfnamefont{R.~L.}\ \bibnamefont{Davidchack}}\ and\
  \bibinfo {author} {\bibfnamefont{B.~B.}\ \bibnamefont{Laird}},\ }%
  \bibfield{journal}{%
  \bibinfo {journal} {J. Chem. Phys.}\ }%
  \textbf{\bibinfo {volume} {118}},\ \bibinfo {pages} {7651} (\bibinfo {year}
  {2003})%
  \bibAnnoteFile{NoStop}{DavLai03}%
\bibitem{HoyAstKar01}%
  \BibitemOpen
  \bibfield{author}{%
  \bibinfo {author} {\bibfnamefont{J.~J.}\ \bibnamefont{Hoyt}}, \bibinfo
  {author} {\bibfnamefont{M.}~\bibnamefont{Asta}},\ and\ \bibinfo {author}
  {\bibfnamefont{A.}~\bibnamefont{Karma}},\ }%
  \bibfield{journal}{%
  \bibinfo {journal} {Phys. Rev. Lett.}\ }%
  \textbf{\bibinfo {volume} {86}},\ \bibinfo {pages} {5530} (\bibinfo {year}
  {2001})%
  \bibAnnoteFile{NoStop}{HoyAstKar01}%
\bibitem{MisAstLi10}%
  \BibitemOpen
  \bibfield{author}{%
  \bibinfo {author} {\bibfnamefont{Y.}~\bibnamefont{Mishin}}, \bibinfo {author}
  {\bibfnamefont{M.}~\bibnamefont{Asta}},\ and\ \bibinfo {author}
  {\bibfnamefont{J.}~\bibnamefont{Li}},\ }%
  \bibfield{journal}{%
  \bibinfo {journal} {Acta Mater.}\ }%
  \textbf{\bibinfo {volume} {58}},\ \bibinfo {pages} {1117} (\bibinfo {year}
  {2010})%
  \bibAnnoteFile{NoStop}{MisAstLi10}%
\bibitem{AngCerLee10}%
  \BibitemOpen
  \bibfield{author}{%
  \bibinfo {author} {\bibfnamefont{S.}~\bibnamefont{{Angioletti-Uberti}}},
  \bibinfo {author} {\bibfnamefont{M.}~\bibnamefont{Ceriotti}}, \bibinfo
  {author} {\bibfnamefont{P.~D.}\ \bibnamefont{Lee}},\ and\ \bibinfo {author}
  {\bibfnamefont{M.~W.}\ \bibnamefont{Finnis}},\ }%
  \bibfield{journal}{%
  \bibinfo {journal} {Phys. Rev. B}\ }%
  \textbf{\bibinfo {volume} {81}},\ \bibinfo {pages} {125416} (\bibinfo {year}
  {2010})%
  \bibAnnoteFile{NoStop}{AngCerLee10}%
\bibitem{Het87}%
  \BibitemOpen
  \bibfield{author}{%
  \bibinfo {author} {\bibfnamefont{J.~H.}\ \bibnamefont{Hetherington}},\ }%
  \bibfield{journal}{%
  \bibinfo {journal} {J. Low Temperature Phys.}\ }%
  \textbf{\bibinfo {volume} {66}},\ \bibinfo {pages} {145} (\bibinfo {year}
  {1987})%
  \bibAnnoteFile{NoStop}{Het87}%
\bibitem{JohPlaViv03}%
  \BibitemOpen
  \bibfield{author}{%
  \bibinfo {author} {\bibfnamefont{R.~S.}\ \bibnamefont{Johal}}, \bibinfo
  {author} {\bibfnamefont{A.}~\bibnamefont{Planes}},\ and\ \bibinfo {author}
  {\bibfnamefont{E.}~\bibnamefont{Vives}},\ }%
  \bibfield{journal}{%
  \bibinfo {journal} {Phys. Rev. E}\ }%
  \textbf{\bibinfo {volume} {68}},\ \bibinfo {pages} {056113} (\bibinfo {year}
  {2003})%
  \bibAnnoteFile{NoStop}{JohPlaViv03}%
\bibitem{TorVal77}%
  \BibitemOpen
  \bibfield{author}{%
  \bibinfo {author} {\bibfnamefont{G.~M.}\ \bibnamefont{Torrie}}\ and\ \bibinfo
  {author} {\bibfnamefont{J.~P.}\ \bibnamefont{Valleau}},\ }%
  \bibfield{journal}{%
  \bibinfo {journal} {J. Comp. Phys.}\ }%
  \textbf{\bibinfo {volume} {23}},\ \bibinfo {pages} {187} (\bibinfo {year}
  {1977})%
  \bibAnnoteFile{NoStop}{TorVal77}%
\bibitem{KasThi05}%
  \BibitemOpen
  \bibfield{author}{%
  \bibinfo {author} {\bibfnamefont{J.}~\bibnamefont{K\"astner}}\ and\ \bibinfo
  {author} {\bibfnamefont{W.}~\bibnamefont{Thiel}},\ }%
  \bibfield{journal}{%
  \bibinfo {journal} {J. Chem. Phys.}\ }%
  \textbf{\bibinfo {volume} {123}},\ \bibinfo {pages} {144104} (\bibinfo {year}
  {2005})%
  \bibAnnoteFile{NoStop}{KasThi05}%
\bibitem{FerSwe88}%
  \BibitemOpen
  \bibfield{author}{%
  \bibinfo {author} {\bibfnamefont{A.~M.}\ \bibnamefont{Ferrenberg}}\ and\
  \bibinfo {author} {\bibfnamefont{R.~H.}\ \bibnamefont{Swendsen}},\ }%
  \bibfield{journal}{%
  \bibinfo {journal} {Phys. Rev. Lett.}\ }%
  \textbf{\bibinfo {volume} {61}},\ \bibinfo {pages} {2635} (\bibinfo {year}
  {1988})%
  \bibAnnoteFile{NoStop}{FerSwe88}%
\bibitem{FerSwe89}%
  \BibitemOpen
  \bibfield{author}{%
  \bibinfo {author} {\bibfnamefont{A.~M.}\ \bibnamefont{Ferrenberg}}\ and\
  \bibinfo {author} {\bibfnamefont{R.~H.}\ \bibnamefont{Swendsen}},\ }%
  \bibfield{journal}{%
  \bibinfo {journal} {Phys. Rev. Lett.}\ }%
  \textbf{\bibinfo {volume} {63}},\ \bibinfo {pages} {1195} (\bibinfo {year}
  {1989})%
  \bibAnnoteFile{NoStop}{FerSwe89}%
\bibitem{Rou95}%
  \BibitemOpen
  \bibfield{author}{%
  \bibinfo {author} {\bibfnamefont{B.}~\bibnamefont{Roux}},\ }%
  \bibfield{journal}{%
  \bibinfo {journal} {Comp. Phys. Comm.}\ }%
  \textbf{\bibinfo {volume} {91}},\ \bibinfo {pages} {275} (\bibinfo {month}
  {Sep.}\ \bibinfo {year} {1995})%
  \bibAnnoteFile{NoStop}{Rou95}%
\bibitem{DicLegLel10}%
  \BibitemOpen
  \bibfield{author}{%
  \bibinfo {author} {\bibfnamefont{B.~M.}\ \bibnamefont{Dickson}}, \bibinfo
  {author} {\bibfnamefont{F.}~\bibnamefont{Legoll}}, \bibinfo {author}
  {\bibfnamefont{T.}~\bibnamefont{Leli\'evre}}, \bibinfo {author}
  {\bibfnamefont{G.}~\bibnamefont{Stoltz}},\ and\ \bibinfo {author}
  {\bibfnamefont{P.}~\bibnamefont{Fleurat-Lessard}},\ }%
  \bibfield{journal}{%
  \bibinfo {journal} {J. Phys. Chem. B}\ }%
  \textbf{\bibinfo {volume} {114}},\ \bibinfo {pages} {5823} (\bibinfo {year}
  {2010})%
  \bibAnnoteFile{NoStop}{DicLegLel10}%
\bibitem{AdjAthRod11}%
  \BibitemOpen
  \bibfield{author}{%
  \bibinfo {author} {\bibfnamefont{G.}~\bibnamefont{Adjanor}}, \bibinfo
  {author} {\bibfnamefont{M.}~\bibnamefont{Ath\`enes}},\ and\ \bibinfo {author}
  {\bibfnamefont{J.~M.}\ \bibnamefont{Rodgers}},\ }%
  \bibfield{journal}{%
  \bibinfo {journal} {J. Chem. Phys.}\ }%
  \textbf{\bibinfo {volume} {135}},\ \bibinfo {pages} {044127} (\bibinfo {year}
  {2011})%
  \bibAnnoteFile{NoStop}{AdjAthRod11}%
\bibitem{SadErhStu12}%
  \BibitemOpen
  \bibfield{author}{%
  \bibinfo {author} {\bibfnamefont{B.}~\bibnamefont{Sadigh}}, \bibinfo {author}
  {\bibfnamefont{P.}~\bibnamefont{Erhart}}, \bibinfo {author}
  {\bibfnamefont{A.}~\bibnamefont{Stukowski}}, \bibinfo {author}
  {\bibfnamefont{A.}~\bibnamefont{Caro}}, \bibinfo {author}
  {\bibfnamefont{E.}~\bibnamefont{Martinez}},\ and\ \bibinfo {author}
  {\bibfnamefont{L.}~\bibnamefont{{Zepeda-Ruiz}}},\ }%
  \bibfield{journal}{%
  \bibinfo {journal} {Phys. Rev. B}\ }%
  \textbf{\bibinfo {volume} {85}},\ \bibinfo {pages} {184203} (\bibinfo {year}
  {2012})%
  \bibAnnoteFile{NoStop}{SadErhStu12}%
\bibitem{Stu10}%
  \BibitemOpen
  \bibfield{author}{%
  \bibinfo {author} {\bibfnamefont{A.}~\bibnamefont{Stukowski}},\ }%
  \bibfield{journal}{%
  \bibinfo {journal} {Model. Simul. Mater. Sci. Eng.}\ }%
  \textbf{\bibinfo {volume} {18}},\ \bibinfo {pages} {015012} (\bibinfo {year}
  {2010})%
  \bibAnnoteFile{NoStop}{Stu10}%
\bibitem{SanDucGra84}%
  \BibitemOpen
  \bibfield{author}{%
  \bibinfo {author} {\bibfnamefont{J.~M.}\ \bibnamefont{Sanchez}}, \bibinfo
  {author} {\bibfnamefont{F.}~\bibnamefont{Ducastelle}},\ and\ \bibinfo
  {author} {\bibfnamefont{D.}~\bibnamefont{Gratias}},\ }%
  \bibfield{journal}{%
  \bibinfo {journal} {Physica}\ }%
  \textbf{\bibinfo {volume} {128}},\ \bibinfo {pages} {334} (\bibinfo {year}
  {1984})%
  \bibAnnoteFile{NoStop}{SanDucGra84}%
\bibitem{AstOzoWoo01}%
  \BibitemOpen
  \bibfield{author}{%
  \bibinfo {author} {\bibfnamefont{M.}~\bibnamefont{Asta}}, \bibinfo {author}
  {\bibfnamefont{V.}~\bibnamefont{Ozolins}},\ and\ \bibinfo {author}
  {\bibfnamefont{C.}~\bibnamefont{Woodward}},\ }%
  \bibfield{journal}{%
  \bibinfo {journal} {JOM}\ }%
  \textbf{\bibinfo {volume} {53}},\ \bibinfo {pages} {16} (\bibinfo {year}
  {2001})%
  \bibAnnoteFile{NoStop}{AstOzoWoo01}%
\bibitem{Note1}%
  \BibitemOpen
  \bibinfo {note} {In general, the two phases that are separated by the
  interface can exhibit size mismatch leading to an additional strain
  contribution to the interface (free) energy.\cite {Mis04, MisAstLi10} This
  contribution can be added to the present approach without further
  complications. In the case of the Fe--Cr system the size mismatch is
  miniscule and the strain contribution is neglible.}%
  \bibAnnoteFile{Stop}{Note1}%
\bibitem{CarCroCar05}%
  \BibitemOpen
  \bibfield{author}{%
  \bibinfo {author} {\bibfnamefont{A.}~\bibnamefont{Caro}}, \bibinfo {author}
  {\bibfnamefont{D.~A.}\ \bibnamefont{Crowson}},\ and\ \bibinfo {author}
  {\bibfnamefont{M.}~\bibnamefont{Caro}},\ }%
  \bibfield{journal}{%
  \bibinfo {journal} {Phys. Rev. Lett.}\ }%
  \textbf{\bibinfo {volume} {95}},\ \bibinfo {pages} {075702} (\bibinfo {year}
  {2005})%
  \bibAnnoteFile{NoStop}{CarCroCar05}%
\bibitem{StuSadErh09}%
  \BibitemOpen
  \bibfield{author}{%
  \bibinfo {author} {\bibfnamefont{A.}~\bibnamefont{Stukowski}}, \bibinfo
  {author} {\bibfnamefont{B.}~\bibnamefont{Sadigh}}, \bibinfo {author}
  {\bibfnamefont{P.}~\bibnamefont{Erhart}},\ and\ \bibinfo {author}
  {\bibfnamefont{A.}~\bibnamefont{Caro}},\ }%
  \bibfield{journal}{%
  \bibinfo {journal} {Model. Simul. Mater. Sci. Eng.}\ }%
  \textbf{\bibinfo {volume} {17}},\ \bibinfo {pages} {075005} (\bibinfo {year}
  {2009})%
  \bibAnnoteFile{NoStop}{StuSadErh09}%
\bibitem{BonErhCar09}%
  \BibitemOpen
  \bibfield{author}{%
  \bibinfo {author} {\bibfnamefont{G.}~\bibnamefont{Bonny}}, \bibinfo {author}
  {\bibfnamefont{P.}~\bibnamefont{Erhart}}, \bibinfo {author}
  {\bibfnamefont{A.}~\bibnamefont{Caro}}, \bibinfo {author}
  {\bibfnamefont{R.~C.}\ \bibnamefont{Pasianot}}, \bibinfo {author}
  {\bibfnamefont{L.}~\bibnamefont{Malerba}},\ and\ \bibinfo {author}
  {\bibfnamefont{M.}~\bibnamefont{Caro}},\ }%
  \bibfield{journal}{%
  \bibinfo {journal} {Model. Simul. Mater. Sci. Eng.}\ }%
  \textbf{\bibinfo {volume} {17}},\ \bibinfo {pages} {025006} (\bibinfo {year}
  {2009})%
  \bibAnnoteFile{NoStop}{BonErhCar09}%
\bibitem{ErhCarCar08}%
  \BibitemOpen
  \bibfield{author}{%
  \bibinfo {author} {\bibfnamefont{P.}~\bibnamefont{Erhart}}, \bibinfo {author}
  {\bibfnamefont{A.}~\bibnamefont{Caro}}, \bibinfo {author}
  {\bibfnamefont{M.}~\bibnamefont{Serrano~de Caro}},\ and\ \bibinfo {author}
  {\bibfnamefont{B.}~\bibnamefont{Sadigh}},\ }%
  \bibfield{journal}{%
  \bibinfo {journal} {Phys. Rev. B}\ }%
  \textbf{\bibinfo {volume} {77}},\ \bibinfo {pages} {134206} (\bibinfo {year}
  {2008})%
  \bibAnnoteFile{NoStop}{ErhCarCar08}%
\bibitem{Note2}%
  \BibitemOpen
  \bibinfo {note} {Volume change trial moves were still included in order to
  allow the system to relax lattice strain.}%
  \bibAnnoteFile{Stop}{Note2}%
\bibitem{ErhMarSad12}%
  \BibitemOpen
  \bibfield{author}{%
  \bibinfo {author} {\bibfnamefont{P.}~\bibnamefont{Erhart}}, \bibinfo {author}
  {\bibfnamefont{J.}~\bibnamefont{Marian}},\ and\ \bibinfo {author}
  {\bibfnamefont{B.}~\bibnamefont{Sadigh}},\ }%
  \enquote{\bibinfo {title} {{Equilibrium structure, shape and orientation
  relations of BCC and 9R Cu-precipitates in Fe from atomistic simulations}},}\
   (\bibinfo {year} {2012}),\ \bibinfo {note} {to be submitted}%
  \bibAnnoteFile{NoStop}{ErhMarSad12}%
\bibitem{Mis04}%
  \BibitemOpen
  \bibfield{author}{%
  \bibinfo {author} {\bibfnamefont{Y.}~\bibnamefont{Mishin}},\ }%
  \bibfield{journal}{%
  \bibinfo {journal} {Acta Mater.}\ }%
  \textbf{\bibinfo {volume} {52}},\ \bibinfo {pages} {1451} (\bibinfo {year}
  {2004})%
  \bibAnnoteFile{NoStop}{Mis04}%
\end{thebibliography}
\end{document}